\documentclass[%
twocolumn,
groupedaddress,
amsmath,amssymb,
aps,
pra,
]{revtex4-2}

\usepackage{graphicx}
\usepackage{amsfonts}
\usepackage{bm}
\usepackage{hyperref,url}
\usepackage{changes}
\usepackage{xcolor}
\usepackage{soul}

\graphicspath{{figures/}}

\begin{document}

\title{Effective skyrmion number for mixed polarization states of light}

\author{Rosario Martínez-Herrero}
\author{Ángel S. Sanz}
\affiliation{Department of Optics, Faculty of Physical Sciences, Universidad Complutense de Madrid, Pza. Ciencias 1, Ciudad Universitaria, 28040 Madrid, Spain}

\begin{abstract}
Polarization skyrmions are usually characterized through a unit Stokes vector field, which defines a map from the transverse plane to the Poincaré (Bloch) sphere and allows one to assign a skyrmion number under appropriate boundary conditions. This description assumes complete polarization. In many optical situations, however, the local polarization state is partially polarized and is therefore described by a spectral polarization matrix rather than by a single Jones vector. We formulate a density-matrix-based effective extension of the skyrmion-number construction for statistically stationary, quasi-monochromatic paraxial fields in terms of the normalized spectral polarization matrix. Its eigenvectors define two orthogonal polarization eigenstates, corresponding to antipodal points on the sphere, while the eigenvalue imbalance is fixed by the local degree of polarization (DoP) $P(\mathbf r)$. Averaging the opposite skyrmion-density contributions of these two eigenpolarizations leads to an effective skyrmion number in which the usual pure-state density is weighted by $P(\mathbf r)$.
The resulting quantity reduces to the conventional skyrmion number for fully polarized fields and assigns zero weight to locally unpolarized regions.
In general, it is not an integer topological invariant; rather, it is an integrated effective measure, constructed from local Stokes parameters, which can be interpreted geometrically as a DoP-weighted signed solid-angle integral of the polarization eigentexture on the Poincar\'e sphere.
We illustrate the construction with a regular analytic family of polarization eigentextures with winding order \(m\) and fixed radial DoP profiles, showing how the effective number depends jointly on the winding order and on the spatial overlap between the DoP and the skyrmion density.
\end{abstract}

\maketitle


\section{\label{sec1} Introduction}

Polarization textures provide a natural setting in which topology enters classical optics. 
When the local polarization state varies across the transverse plane, the associated Stokes vector may trace a nontrivial pattern on the Poincar\'e, or equivalently Bloch, sphere.
In particular, polarization skyrmions are commonly described as maps from the transverse coordinate 
${\bf r}=(x,y)$ to a unit vector on this sphere, with the corresponding skyrmion number measuring the signed wrapping of the sphere under suitable boundary conditions 
\cite{gao:PRA:2020,sonja:LaserPhotonRev:2023,zayats:NatPhotonics:2024,cheng:AdvOptPhot:2025,forbes2026AdvPhoton,MataCervera2026}.

This construction is geometrically transparent, but it relies on an important assumption: the field is taken to be fully polarized at every point. 
Under this condition the local polarization state is represented by a Jones vector, or equivalently by a unit Stokes vector, and the usual skyrmion number is the degree of a map from the compactified transverse plane to the Poincar\'e sphere. 
The situation is less direct for partially polarized light. 
In that case the Stokes vector no longer lies on the surface of the sphere, but inside it, with a length given by the local degree of polarization (DoP). 
Thus the same optical field carries two distinct pieces of information: a polarization direction, when it is defined, and a local purity that quantifies how well defined that direction is.

Nonuniform partially polarized fields are not exceptional. They arise naturally from imperfect state preparation, incoherent superpositions, partial spatial or temporal coherence, scattering, or propagation through random or complex media. For such fields, a description in terms of a single Jones vector is insufficient. The standard object is instead the cross-spectral density matrix, and locally the spectral polarization matrix, which contains the second-order polarization information of a statistically stationary, quasi-monochromatic field \cite{mandelwolf-bk,wolf-bk-2}. Recent studies of partially coherent and stochastic optical skyrmions have further highlighted the need to distinguish the geometry of the polarization texture from coherence- or purity-dependent changes in its observable structure \cite{Chen:OL:2026,Liu:arXiv:2026}. After normalization, this matrix has the same structure as a two-level density matrix: its eigenvectors define two orthogonal polarization eigenstates, while its eigenvalues give their statistical weights.

This density-matrix structure provides a useful starting point for revisiting polarization skyrmions in mixed polarization states.
At each point, the local polarization matrix can be viewed as an incoherent mixture of two antipodal pure polarization states on the Poincar\'e sphere. 
The imbalance between their weights is fixed by the DoP. 
Since the skyrmion density changes sign when the Bloch vector is inverted, the two eigenpolarization contributions do not add equally; their net contribution is controlled by this eigenvalue imbalance.

The central issue is therefore not whether a partially polarized field possesses an integer topological invariant of the same kind as a pure polarization texture. 
In general it does not. 
Rather, the relevant question is how to define a quantity that reduces to the usual skyrmion number in the fully polarized limit, gives no contribution from locally unpolarized regions, and remains expressed in terms of measurable polarization quantities. 
This is the point of departure for the effective skyrmion number considered below.

The quantity introduced here is best understood as a measure of effective skyrmionic content, not as a new homotopy invariant. It quantifies the skyrmion density of the polarization eigentexture after each local contribution is weighted by the polarized fraction of the field. This distinction is relevant because two fields may have the same eigentexture, and hence the same conventional skyrmion number, while differing in how their polarized fractions are distributed across the beam. The conventional skyrmion number does not distinguish between such fields, whereas the effective number introduced here does.
Although recent approaches have addressed skyrmionic structures in partially coherent optical fields and mixed quantum states of light \cite{Chen:OL:2026,Liu:arXiv:2026,Kam:arXiv:2026},
to the best of our knowledge the specific DoP-weighted construction considered here has not been explicitly developed in the optical-skyrmion literature.

In this work we formulate such a quantity by taking the local spectral polarization matrix as the basic object. Section~\ref{sec2} reviews its mixed-state structure and separates the polarization eigentexture from the local DoP.
Section~\ref{sec3} introduces the corresponding effective skyrmion number, relates the DoP weighting to the spectral decomposition of the polarization matrix, and clarifies its geometric meaning. Section~\ref{sec4} applies the construction to a regular analytic family of polarization eigentextures with winding order \(m\) while keeping two radial DoP profiles fixed. This allows us to distinguish the influence of winding order from that of the spatial distribution of the polarized fraction.
The conclusions are given in Sec.~\ref{sec5}.


\section{\label{sec2} Polarization matrix and mixed-state structure}

We start from the standard local description of partially polarized paraxial fields. Consider a statistically stationary, quasi-monochromatic electromagnetic field whose transverse components are denoted by $E_i({\bf r})$, with ${\bf r}=(x,y)$. Its second-order coherence and polarization properties are described by the cross-spectral density matrix
\begin{equation}
\hat{\mathcal{W}}_{ij}({\bf r}_1,{\bf r}_2) = \left\langle E_i({\bf r}_1)E_j^*({\bf r}_2) \right\rangle ,
\label{eq:CSDM}
\end{equation}
where $\langle\ \bullet\ \rangle$ denotes an ensemble average. The local polarization state is obtained by evaluating the cross-spectral density matrix at coincident points,
\begin{equation}
\hat{\mathcal{J}}({\bf r}) \equiv \hat{\mathcal{W}}({\bf r},{\bf r}) .
\label{eq:Jlocal}
\end{equation}
The matrix $\hat{\mathcal{J}}({\bf r})$ is Hermitian and positive semidefinite. Its trace gives the local spectral density, whereas its normalized form contains the local polarization state.
Different spatial-coherence properties may lead to the same local polarization matrix
\cite{MartinezHerrero:OptLett:2007}; this does not affect the present construction, which is based only on local polarization information.

Using the Pauli matrices
\begin{equation}
\hat{\sigma}_1=
\begin{pmatrix}
0&1\\
1&0
\end{pmatrix},
\quad
\hat{\sigma}_2=
\begin{pmatrix}
0&-i\\
i&0
\end{pmatrix},
\quad
\hat{\sigma}_3=
\begin{pmatrix}
1&0\\
0&-1
\end{pmatrix},
\label{eq:Pauli}
\end{equation}
with \(\hat{\boldsymbol{\sigma}} =(\hat{\sigma}_1,\hat{\sigma}_2,\hat{\sigma}_3)\),
together with $\hat{\sigma}_0=\hat{I}_2$, the spectral polarization matrix can be written in the Stokes representation \cite{schmieder:JOSA:1969,brosseau-bk,MartinezHerrero-bk} as
\begin{equation}
\hat{\mathcal{J}}({\bf r})
= \frac{S_0 ({\bf r})}{2}
\left[ \hat{I}_2 + {\bf s} ({\bf r}) \cdot \hat{\boldsymbol{\sigma}} \right] .
\label{eq:JPauli}
\end{equation}
With this convention, the normalized Stokes vector is
\begin{equation}
{\bf s}({\bf r}) \equiv \frac{1}{S_0({\bf r})} \begin{pmatrix} S_1({\bf r}) \\ S_2({\bf r}) \\ S_3({\bf r}) \end{pmatrix} ,
\label{eq:svec}
\end{equation}
whose length corresponds to the local DoP,
\begin{equation}
 P({\bf r}) \equiv \| {\bf s}({\bf r})\| = \frac{\sqrt{S_1^2({\bf r})+S_2^2({\bf r})+S_3^2({\bf r})}}{S_0({\bf r})}.
\label{eq:DoP}
\end{equation}
Thus $0\le P({\bf r})\le1$.
A fully polarized state corresponds to $P({\bf r})=1$, whereas $P({\bf r})=0$ corresponds to a locally unpolarized state \cite{mandelwolf-bk,wolf-bk-2}.
Equivalently, Eq.~\eqref{eq:JPauli} can be written explicitly in matrix form as
\begin{equation}
\hat{\mathcal{J}}({\bf r}) = \frac12
\begin{pmatrix}
S_0+S_3 & S_1-iS_2\\
S_1+iS_2 & S_0-S_3
\end{pmatrix} ,
\label{eq:Jstokes}
\end{equation}
where all Stokes parameters are evaluated at \({\bf r}\).

The density-matrix structure becomes explicit after normalizing the polarization matrix by its trace, \(S_0({\bf r})=\operatorname{Tr}\hat J({\bf r})\):
\begin{equation}
\hat{\rho}({\bf r}) = \frac{\hat{\mathcal{J}}({\bf r})}{S_0({\bf r})}.
\label{eq:rhoDef}
\end{equation}
Since $\hat{\mathcal{J}}$ is Hermitian and positive semidefinite, it admits the spectral decomposition
\begin{equation}
\hat{\mathcal{J}}({\bf r}) = \lambda_+({\bf r})
|u_+({\bf r})\rangle
\langle u_+({\bf r})|
+
\lambda_-({\bf r})
|u_-({\bf r})\rangle
\langle u_-({\bf r})| ,
\label{eq:spectral}
\end{equation}
where $\lambda_+({\bf r})\ge\lambda_-({\bf r})\ge0$ and the eigenvectors are orthonormal.
The eigenvalues are fixed by $S_0$ and by the DoP,
\begin{equation}
\lambda_\pm ({\bf r}) = \frac{S_0({\bf r})}{2} \Big[ 1 \pm \| {\bf s} ({\bf r}) \| \Big]
 = \frac{S_0({\bf r})}{2} \Big[ 1 \pm P({\bf r}) \Big] .
\label{eq:lambdas}
\end{equation}
Thus partial polarization is locally equivalent to an incoherent mixture of two orthogonal fully polarized eigenstates, with an imbalance determined by $P({\bf r})$.

For \(P({\bf r})\neq0\), the dominant eigenpolarization defines a direction on the Poincar\'e sphere. Introducing the corresponding spherical angles \(\theta\) and \(\phi\), the Stokes parameters can be written as
\begin{equation}
\begin{aligned}
 S_1({\bf r})&=S_0({\bf r})P({\bf r})\sin\theta\cos\phi,\\
 S_2({\bf r})&=S_0({\bf r})P({\bf r})\sin\theta\sin\phi,\\
 S_3({\bf r})&=S_0({\bf r})P({\bf r})\cos\theta.
\end{aligned}
\label{eq:StokesSphericalParametrization}
\end{equation}
A convenient choice of normalized eigenvectors is
\begin{equation}
|u_+\rangle =
\begin{pmatrix}
\cos(\theta/2)\\
e^{i\phi}\sin(\theta/2)
\end{pmatrix},
\quad
|u_-\rangle =
\begin{pmatrix}
-e^{-i\phi}\sin(\theta/2)\\
\cos(\theta/2)
\end{pmatrix}.
\label{eq:evecs}
\end{equation}
These eigenvectors correspond to the antipodal unit Bloch vectors \(\pm\hat{\bf n}_0({\bf r})\), with
\begin{equation}
\hat{\bf n}_0({\bf r})
=
\bigl(
\sin\theta\cos\phi,\,
\sin\theta\sin\phi,\,
\cos\theta
\bigr) ,
\label{eq:n0}
\end{equation}
and satisfy
\begin{equation}
 \hat{\bf n}_0 \cdot \hat{\boldsymbol{\sigma}} |u_\pm \rangle = \pm |u_\pm \rangle .
\end{equation}
The normalized polarization matrix then takes the Bloch form
\begin{equation}
\hat{\rho}({\bf r}) = \frac12 \left[ \hat{I}_2 + P({\bf r}) \hat{\bf n}_0({\bf r})\cdot\hat{\boldsymbol{\sigma}} \right].
\label{eq:rhoBloch}
\end{equation}
Equation~\eqref{eq:rhoBloch} has the same form as the usual Bloch representation of a mixed two-level state, or qubit density matrix. 
In the present optical context, the corresponding Bloch sphere is identified with the Poincar\'e sphere of polarization states; we use the two terms interchangeably when referring to this geometric state space.
It separates the two ingredients that will enter the topological construction below: the eigentexture $\hat{\bf n}_0({\bf r})$, which gives the direction on the Poincar\'e sphere, and the DoP $P({\bf r})$, which gives the local purity of the state. When $P=1$, the state lies on the surface of the sphere and is fully polarized. When $0<P<1$, the direction remains defined by the eigenpolarization, but the Bloch vector has length $P$.
At points where \(P=0\), the polarization direction is locally undefined.

Equivalently, the local state can be written as the sum of a fully polarized component and an unpolarized background,
\begin{equation}
\hat{\mathcal{J}}({\bf r}) = S_0({\bf r}) P({\bf r}) |E_p({\bf r})\rangle\langle E_p({\bf r})| + \frac{S_0({\bf r})}{2} \bigl[1-P({\bf r})\bigr]\hat{I}_2 ,
 \label{eq:Jexample}
\end{equation}
where \(|E_p({\bf r})\rangle=|u_+({\bf r})\rangle\) wherever \(P({\bf r})>0\).
This form follows directly from the spectral decomposition of the positive-semidefinite polarization matrix \cite{graybill1983matrices,brosseau-bk}.
The first term carries the polarized part of the field, whereas the second term is locally unpolarized and carries no direction on the Poincar\'e sphere.


\section{\label{sec3} Effective skyrmion number for mixed polarization states}

For a fully polarized field, the local polarization state is represented by a unit vector
$\hat{\bf n}_0({\bf r})$ on the Poincar\'e sphere. If this vector field approaches a
constant value at infinity, the transverse plane can be compactified to a sphere and
$\hat{\bf n}_0({\bf r})$ defines a map $S^2\to S^2$. The corresponding skyrmion number (topological charge) is
\begin{equation}
N = \frac{1}{4\pi} \iint_{\mathbb R^2} q(\hat{\bf n}_0)
\,dx\,dy ,
\label{eq:Npure}
\end{equation}
where
\begin{equation}
q(\hat{\bf n}_0) \equiv
\hat{\bf n}_0\cdot \left( \partial_x\hat{\bf n}_0 \times \partial_y\hat{\bf n}_0 \right)
\label{eq:qdef}
\end{equation}
denotes the unnormalized skyrmion number density (topological charge density) for a pure texture.
The integrand is the signed solid-angle density swept on the Poincar\'e sphere by an
infinitesimal area element of the transverse plane. Under the above boundary condition,
$N$ is the degree of the map and is an integer topological invariant
\cite{han-bk,sonja:LaserPhotonRev:2023,belavin:JETPLett:1975}. If the boundary
condition is not satisfied, or if the integration is restricted to a finite observation
window, Eq.~\eqref{eq:Npure} remains well defined but need not yield an integer
\cite{sonja:PRSA:2024}.

For a partially polarized field, the normalized polarization matrix has the Bloch form
given in Eq.~\eqref{eq:rhoBloch}. The local state is then represented by the Bloch vector
\begin{equation}
{\bf s}({\bf r}) = P({\bf r})\hat{\bf n}_0({\bf r}),
\label{eq:blochVector}
\end{equation}
whose length is the DoP. Thus the geometry of the eigentexture is
still described by the unit vector $\hat{\bf n}_0({\bf r})$, whereas the factor
$P({\bf r})$ measures how pure, or how well polarized, the local state is.
In this case the state generally lies inside the Poincar\'e ball and the integer topological interpretation of Eq.~\eqref{eq:Npure} is no longer available in terms of the mixed state itself.

The spectral decomposition of the local polarization matrix provides a density-matrix route to an effective density.
The construction below should therefore be understood as a density-matrix-based effective quantity, not as a claim of a unique topological invariant for mixed polarization states.
The normalized eigenvalues are
\begin{equation}
\mu_\pm({\bf r}) \equiv \frac{\lambda_\pm({\bf r})}{S_0({\bf r})} = \frac{1 \pm P({\bf r})}{2} .
\label{eq:muab}
\end{equation}
The two eigenpolarizations correspond to antipodal directions $\pm\hat{\bf n}_0({\bf r})$ on the Poincar\'e sphere, as mentioned above.
Since the density defined in Eq.~\eqref{eq:qdef} changes sign under inversion of the Bloch vector, $q(-\hat{\bf n}_0)=-q(\hat{\bf n}_0)$, the density obtained by averaging over the two eigenpolarization contributions is
\begin{eqnarray}
q_{\rm mix}({\bf r}) & = & \mu_+({\bf r})\,q(\hat{\bf n}_0) + \mu_-({\bf r})\,q(-\hat{\bf n}_0) \nonumber \\
 & = & \left[ \mu_+({\bf r}) - \mu_-({\bf r}) \right] q(\hat{\bf n}_0) = P({\bf r})\,q(\hat{\bf n}_0).
\label{eq:qmix}
\end{eqnarray}
Thus the linear factor \(P({\bf r})\), which is also the length of the Bloch vector, follows from the local eigenvalue imbalance between the two antipodal pure-state contributions.
In this sense, the weighting is not imposed independently at the level of the integral; it results from combining the two eigenpolarization contributions linearly with their statistical weights. Other nonlinear scalar weights may be introduced as alternative diagnostics, but they do not correspond to this spectral average.

We therefore define the effective skyrmion number of a partially polarized field as
\begin{equation}
N_{\rm eff} \equiv \frac{1}{4\pi} \iint_{\mathbb R^2} P({\bf r}) q(\hat{\bf n}_0)
\,dx\,dy .
\label{eq:Neff}
\end{equation}
The geometric meaning of Eq.~\eqref{eq:Neff} is inherited from the pure-state eigentexture. The quantity \[ q(\hat{\bf n}_0)\,dx\,dy \] is the oriented solid-angle element swept by \(\hat{\bf n}_0({\bf r})\) on the Poincar\'e sphere. Multiplication by \(P({\bf r})\) weights this local solid-angle contribution by the polarized fraction of the field. Thus, \(N_{\rm eff}\) can be interpreted as a DoP-weighted signed solid-angle integral of the polarization eigentexture.

This interpretation should not be confused with the geometrical area of the surface traced by the nonunit Stokes vector in Eq.~\eqref{eq:blochVector} inside the Poincar\'e ball.
Moreover, it also differs from applying the cubic pure-state expression directly to this vector.
Note that, from Eq.~\eqref{eq:blochVector}, we readily obtain
\begin{equation}
 \partial_i{\bf s} = (\partial_iP)\hat{\bf n}_0 + P\partial_i\hat{\bf n}_0, \qquad i=x,y ,
 \label{eq:StokesVectorDerivative}
\end{equation}
which leads to
\begin{equation}
 {\bf s}\cdot \left( \partial_x{\bf s} \times \partial_y{\bf s} \right) = P^3q(\hat{\bf n}_0) .
 \label{eq:NonunitCubicDensity}
\end{equation}
Thus, neither the cubic expression evaluated on the nonunit Stokes vector nor the geometrical area of the surface traced by \({\bf s}({\bf r})\) inside the Poincar\'e ball coincides with the spectral average in Eq.~\eqref{eq:Neff}.
The latter area would involve
\begin{equation}
 \left|\partial_x{\bf s}\times\partial_y{\bf s}\right|\,dx\,dy\
\end{equation}
and would generally contain spatial derivatives of \(P({\bf r})\).
It therefore represents a different geometric construction from the spectral average considered here.

Experimentally, all quantities entering Eq.~\eqref{eq:Neff} can be obtained from spatially resolved Stokes measurements. The measured Stokes maps determine both the local DoP and, wherever \(P({\bf r})>0\), the eigentexture direction \(\hat{\bf n}_0({\bf r})={\bf s}({\bf r})/P({\bf r})\). The weighted density can then be evaluated numerically over the observation plane. At points where \(P({\bf r})=0\), the eigenvalues of the polarization matrix are degenerate and no unique polarization direction is defined; such points carry zero weight in Eq.~\eqref{eq:Neff}, provided that the weighted density has a regular limit. As in conventional experimental determinations of optical skyrmion numbers, suitable spatial regularization or a DoP threshold may be required in weakly polarized regions, where the reconstructed eigentexture direction becomes sensitive to measurement noise \cite{sonja:LaserPhotonRev:2023,forbes2026AdvPhoton,Chen:OL:2026}.

Equation~\eqref{eq:Neff} reduces to Eq.~\eqref{eq:Npure} in the fully polarized limit \(P({\bf r})=1\). In general, however, \(N_{\rm eff}\) is not an integer and should not be interpreted as a homotopy invariant.
This distinction is important because mixed-state geometry and topology can be treated in other ways, for example through interferometric mixed-state geometric phases, Uhlmann holonomy, or recent coherence--Stokes constructions that encode skyrmionic topology directly in a mixed-state density matrix \cite{sjoqvist:PRL:2000,uhlmann1986parallel,Kam:arXiv:2026}. 
The present construction is instead a local-polarization, Stokes-based quantity derived from the spectral decomposition of the local polarization matrix.
It integrates the skyrmion density of the polarization eigentexture, with each local contribution weighted by the local DoP.

A useful limiting case is obtained when the DoP is uniform,
$P({\bf r})=P_0$. Then Eq.~\eqref{eq:Neff} gives
\begin{equation}
N_{\rm eff}=P_0 N .
\label{eq:NeffConst}
\end{equation}
Uniform depolarization therefore rescales the skyrmion number of the underlying
eigentexture without modifying its geometry. For nonuniform $P({\bf r})$, the effective
number depends on the spatial overlap between the DoP and the local
skyrmion density.
This distinction is illustrated in the order-\(m\) analysis below.


\section{\label{sec4} Analytic examples with polarization eigentextures of winding order \(m\)}

We now apply Eq.~\eqref{eq:Neff} to a cylindrically symmetric Gaussian model.
The fully polarized component generates a regular family of polarization eigentextures with positive integer winding order \(m\), while the same radial DoP profiles are retained for every value of \(m\).
This allows us to distinguish the influence of the winding order from that of the spatial distribution of the polarized fraction.


\subsection{\label{sec4A} Regular order-\texorpdfstring{$m$}{m} eigentexture}

Let $(r,\beta)$ denote polar coordinates in the transverse plane, and consider the Gaussian spectral-density envelope
\begin{equation}
I(r) = I_0\exp\left(-\frac{2r^2}{w_0^2}\right).
\label{eq:GaussianEnvelope}
\end{equation}
We set $S_0({\bf r})=I(r)$ in Eq.~\eqref{eq:Jexample} and specify the fully polarized component through the normalized Jones state
\begin{equation}
|E_p^{(m)}\rangle = \frac{1}{\sqrt{1+|\alpha_m|^2}} \begin{pmatrix} 1 \\ \alpha_m \end{pmatrix},
\label{eq:EpOrderM}
\end{equation}
where
\begin{equation}
\alpha_m(\xi,\beta) = \xi^m e^{i(m\beta+\beta_0)},
\quad \xi=\frac{r}{r_0}, \quad r_0=\frac{w_0}{\sqrt{2}} .
\label{eq:alphaOrderM}
\end{equation}
Here $r_0$ sets the transverse scale of the texture, $\beta_0$ is a constant helicity angle, and $m=1,2,\ldots$ is the winding order.
The radial factor $\xi^m$ ensures regularity at the beam center, since
\begin{equation}
\xi^m e^{im\beta} = \left(\frac{x+iy}{r_0}\right)^m .
\label{eq:regularityOrderM}
\end{equation}
This expression is single valued and regular at $r=0$ for every positive integer $m$.

Let us define
\begin{equation}
 \varphi_m \equiv m\beta + \beta_0 ,
 \label{eq:PhiOrderM}
\end{equation}
and write the Jones state in Eq.~\eqref{eq:EpOrderM} as \((a,b)^T\), so that its Stokes-vector components are
\begin{equation}
 \begin{split}
 \hat n_{m,1} = 2\operatorname{Re}(a^*b) , \\
 \hat n_{m,2} = 2\operatorname{Im}(a^*b) , \\
 \hat n_{m,3}=|a|^2-|b|^2 .
 \end{split}
\end{equation}
The corresponding unit Stokes vector is therefore
\begin{eqnarray}
\hat{\bf n}_m(\xi,\beta) & = & \frac{1}{1 + |\alpha_m|^2}
 \begin{pmatrix} 2\operatorname{Re}(\alpha_m) \\ 2\operatorname{Im}(\alpha_m) \\ 1-|\alpha_m|^2 \end{pmatrix} \nonumber \\
 & = & \frac{1}{1+\xi^{2m}}
\begin{pmatrix} 2\xi^m\cos\varphi_m \\ 2\xi^m\sin\varphi_m \\ 1-\xi^{2m} \end{pmatrix} ,
\label{eq:nOrderM}
\end{eqnarray}
satisfying $\hat{\bf n}_m(0)=(0,0,1)$ and $\hat{\bf n}_m(\infty)=(0,0,-1)$.
As \(\xi\) increases from zero to infinity, the eigentexture moves from the north pole to the south pole of the Poincar\'e sphere.
During one complete circuit around the beam axis, \(\beta:0\to2\pi\), the angle \(\varphi_m\) changes by \(2\pi m\), and the eigentexture winds \(m\) times around the \(S_3\) axis.
The parameter \(\beta_0\) produces a rigid rotation about the \(S_3\) axis without modifying the skyrmion number.
Equation~\eqref{eq:nOrderM} therefore defines a regular stereographic family of order $m$ \cite{franke-arnold:AdvPhotonRes:2023}.
Recently, the structural stability and singularity splitting of high-order polarization textures under perturbations have been analyzed \cite{MataCervera2026}.

Substitution of Eq.~\eqref{eq:nOrderM} into the definition in Eq.~\eqref{eq:qdef} gives
\begin{equation}
q_m(\xi)
=
\hat{\bf n}_m\cdot
\left(
\partial_x\hat{\bf n}_m
\times
\partial_y\hat{\bf n}_m
\right)
=
\frac{4m^2}{r_0^2}
\frac{\xi^{2(m-1)}}
{\left(1+\xi^{2m}\right)^2}.
\label{eq:densityOrderM}
\end{equation}
Substitution into Eq.~\eqref{eq:Npure} gives
\begin{equation}
N_m
=
2m^2
\int_0^\infty
\frac{\xi^{2m-1}}
{\left(1+\xi^{2m}\right)^2}
\,d\xi
=
m.
\label{eq:NOrderM}
\end{equation}
Thus, the family provides regular polarization eigentextures with positive integer skyrmion number $N_m=m$.


\subsection{Fixed radial DoP profiles}

\begin{figure}[!t]
\centering
\includegraphics[width=\columnwidth]{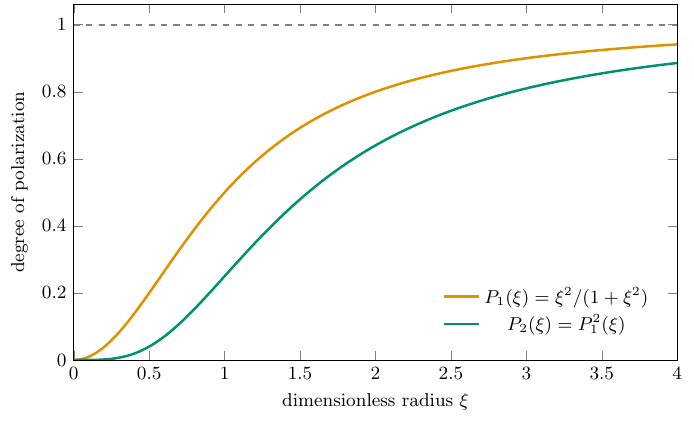}
\caption{\label{fig:DoPprofiles}
Fixed radial DoP profiles $P_1(\xi)$ and $P_2(\xi)$, used for every winding order \(m\). Both profiles increase monotonically from zero to the fully polarized limit $P(\infty)=1$, while \(P_2(\xi)<P_1(\xi)\) for \(0<\xi<\infty\).
}
\end{figure}

To compare different winding orders $m$ under the same partial-polarization conditions, we keep the following two radial DoP profiles fixed for all values of $m$:
\begin{equation}
P_1(\xi) = \frac{\xi^2}{1+\xi^2},
\label{eq:P1}
\end{equation}
and
\begin{equation}
P_2(\xi) \equiv P_1^2(\xi) = \left( \frac{\xi^2}{1+\xi^2} \right)^2.
\label{eq:P2}
\end{equation}
Both profiles vanish at the beam center and approach unity for large $\xi$, although $P_2(\xi)<P_1(\xi)$ for every finite $\xi>0$.
These two profiles are shown in Fig.~\ref{fig:DoPprofiles}.
Note that, because neither profile depends on \(m\), any change in the normalized effective number with the winding order arises from the redistribution of the skyrmion density rather than from a change in the DoP profile.

\subsection{Effective skyrmion number and winding order}

For an arbitrary radial DoP profile $P(\xi)$, Eqs.~\eqref{eq:Neff} and \eqref{eq:densityOrderM} give
\begin{equation}
 N_{\rm eff}^{(m)}[P] = 2m^2 \int_0^\infty P(\xi) \frac{\xi^{2m-1}}{\left(1+\xi^{2m}\right)^2} \, d\xi .
\label{eq:NeffOrderM}
\end{equation}
The two profiles in Eqs.~\eqref{eq:P1} and \eqref{eq:P2} illustrate qualitatively different behaviors.
To simplify notation below, we will use the definitions $N_{{\rm eff},1}^{(m)} \equiv N_{\rm eff}^{(m)}[P_1]$ and $N_{{\rm eff},2}^{(m)} \equiv N_{\rm eff}^{(m)}[P_2]$.

Under the transformation \(\xi\mapsto1/\xi\), the radial measure in Eq.~\eqref{eq:NeffOrderM} is invariant, whereas \(P_1\) transforms into \(1-P_1\). Adding the original and transformed integrals therefore gives
\begin{equation}
 N_{{\rm eff},1}^{(m)} = 2m^2 \int_0^\infty \frac{\xi^{2m+1}}{(1+\xi^2)\left(1+\xi^{2m}\right)^2} \,d\xi = \frac{m}{2}.
\label{eq:NeffP1}
\end{equation}
Thus, $P_1$ assigns exactly one half of the pure skyrmion number to every winding order, even though the radial distribution of the skyrmion density changes with $m$.
In particular,
\begin{equation}
N_{{\rm eff},1}^{(1)}=\frac12,
\qquad
N_{{\rm eff},1}^{(2)}=1,
\qquad
N_{{\rm eff},1}^{(3)}=\frac32.
\label{eq:NeffP1FirstOrders}
\end{equation}

For the second profile,
\begin{equation}
 N_{{\rm eff},2}^{(m)} = 2m^2 \int_0^\infty \frac{\xi^{2m+3}}{(1+\xi^2)^2 \left(1+\xi^{2m}\right)^2} \,d\xi .
\label{eq:NeffP2}
\end{equation}
Unlike Eq.~\eqref{eq:NeffP1}, this expression does not reduce to a fixed fraction of $m$.
The first three values are
\begin{equation}
N_{{\rm eff},2}^{(1)} = \frac13,
\quad
N_{{\rm eff},2}^{(2)} = \frac{\pi}{2}-1 \simeq 0.5708,
\quad
N_{{\rm eff},2}^{(3)} \simeq 0.8061.
\label{eq:NeffP2FirstOrders}
\end{equation}
In particular, the order-two value is not twice the corresponding degree-one result.
The reason is that increasing $m$ changes not only the winding number but also the radial distribution of the skyrmion density sampled by the fixed DoP profile.

Although Eq.~\eqref{eq:NeffP2} does not yield a simple linear rescaling for arbitrary $m$, its behavior can be characterized analytically.
Using \(P_2=P_1^2=(P_1-1/2)^2+P_1-1/4\), together with Eqs.~\eqref{eq:NOrderM} and \eqref{eq:NeffP1}, Eq.~\eqref{eq:NeffP2} can be written as
\begin{equation}
N_{{\rm eff},2}^{(m)} = \frac{m}{4} + \delta_m,
\label{eq:NeffP2Decomposition}
\end{equation}
where
\begin{equation}
\delta_m \equiv 2m^2
\int_0^\infty
\left[
P_1(\xi)-\frac12
\right]^2
\frac{\xi^{2m-1}}
{\left(1+\xi^{2m}\right)^2}
\,d\xi.
\label{eq:DeltaM}
\end{equation}
Consequently,
\begin{equation}
\frac{m}{4}
<
N_{{\rm eff},2}^{(m)}
<
\frac{m}{2}
\label{eq:NeffP2Bounds}
\end{equation}
for every finite positive integer $m$.
The lower bound follows from \(\delta_m>0\), whereas the upper bound follows from \(P_2(\xi)<P_1(\xi)\) for \(0<\xi<\infty\).

The large-\(m\) behavior can be obtained more generally by introducing \(t=\xi^{2m}\) in Eq.~\eqref{eq:NeffOrderM}, which gives
\begin{equation}
 \frac{N_{\rm eff}^{(m)}[P]}{m} = \int_0^\infty \frac{ P\left(t^{1/(2m)}\right) }{ (1+t)^2 } \,dt.
 \label{eq:NeffLargeMGeneral}
\end{equation}
For any bounded radial profile that is continuous at \(\xi=1\), it follows that
\begin{equation}
 \lim_{m\to\infty} \frac{N_{\rm eff}^{(m)}[P]}{m} = P(1).
 \label{eq:NeffLargeMLimit}
\end{equation}
In particular, \(P_2(1)=1/4\), and therefore
\begin{equation}
\lim_{m\to\infty}
\frac{N_{{\rm eff},2}^{(m)}}{m} = \frac14.
\label{eq:NeffP2Limit}
\end{equation}
This result reflects the progressive localization of the normalized radial skyrmion-number distribution around \(\xi=1\) as the winding order increases.

\begin{figure}[!t]
\centering
\includegraphics[width=\columnwidth]{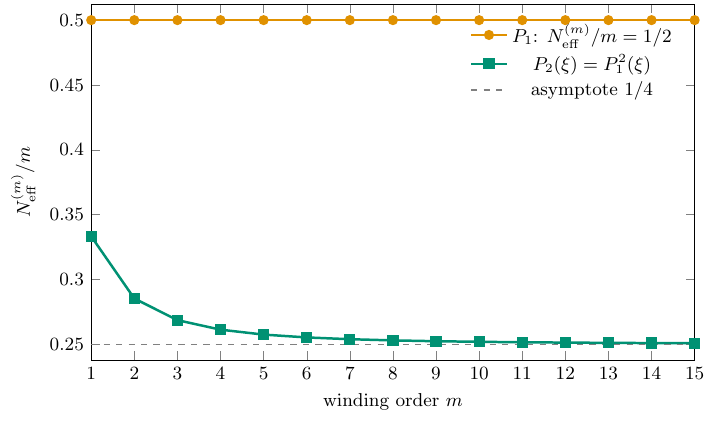}
\caption{\label{fig:NeffOrderM}
Normalized effective skyrmion number $N_{\rm eff}^{(m)}/m$ as a function of the positive integer winding order $m$ for the fixed DoP profiles $P_1$ and $P_2$.
For $P_1$, the normalized value is exactly $1/2$ for every $m$.
For \(P_2\), numerical evaluation of Eq.~\eqref{eq:NeffP2} approaches the general large-\(m\) limit \(P_2(1)=1/4\), shown by the horizontal dashed line.
}
\end{figure}

Figure~\ref{fig:NeffOrderM} summarizes the dependence on the winding order.
For \(P_1\), the normalized effective number remains exactly equal to \(1/2\).
For \(P_2\), it decreases toward the asymptotic value \(1/4\).
The comparison shows that \(N_{\rm eff}\) depends jointly on the winding order and on the spatial overlap between the fixed DoP profile and the order-dependent skyrmion-density distribution.
A complementary comparison in which the eigentexture is kept fixed while the polarized fraction is redistributed radially is given in Appendix~\ref{app:complementary_profiles}.

\section{\label{sec5} Conclusions}

We have examined how the usual skyrmion-number construction for polarization textures can be extended to optical fields whose local polarization state is mixed. The appropriate local description is provided by the spectral polarization matrix. After normalization, this matrix has the form of a two-level density matrix: its eigenvectors define two orthogonal polarization eigenstates associated with antipodal directions on the Poincar\'e sphere, while their eigenvalue imbalance is fixed by the local DoP.

This structure leads to an effective skyrmion number in which the pure-state skyrmion density of the polarization eigentexture is weighted by the local DoP \(P({\bf r})\). The linear weighting is not introduced as an independent assumption; it follows from combining the opposite skyrmion-density contributions of the two antipodal eigenpolarizations with their statistical weights. Other nonlinear scalar weights may be introduced as alternative diagnostics, but they do not correspond to this spectral average.

The resulting effective number is generally not an integer and should not be interpreted as a homotopy invariant of the mixed state. Geometrically, it is a DoP-weighted integral of the signed solid-angle density swept by the polarization eigentexture on the Poincar\'e sphere. It is therefore distinct from the geometrical area of the surface traced by the nonunit Stokes vector inside the Poincar\'e ball. In this sense, the construction preserves the spherical geometry of the eigentexture while accounting locally for the polarized fraction of the field.

The analytic Gaussian model makes explicit how the effective number depends jointly on the winding order and on the spatial distribution of the DoP. For the regular order-\(m\) family, the conventional skyrmion number is \(N_m=m\). The fixed profile \(P_1(\xi)=\xi^2/(1+\xi^2)\) gives \(N_{{\rm eff},1}^{(m)}=m/2\) for every positive integer \(m\). For the second profile, \(P_2(\xi)=P_1^2(\xi)\), the effective number is not a simple rescaling of its degree-one value. More generally, for a bounded radial DoP profile that is continuous at \(\xi=1\), the normalized effective number approaches \(P(1)\) as the winding order increases. In particular, \(P_2(1)=1/4\), so that \(N_{{\rm eff},2}^{(m)}/m\) approaches \(1/4\). This behavior results from the progressive localization of the normalized radial skyrmion-density distribution around \(\xi=1\). It does not represent a change in the topological winding itself, but rather reflects how the polarized fraction weights the local contributions to the effective number.
The complementary-profile comparison in Appendix~\ref{app:complementary_profiles} further illustrates this sensitivity by redistributing the DoP while leaving the polarization eigentexture unchanged.

The formulation presented here is therefore best viewed as a density-matrix-based assignment of effective skyrmionic content to partially polarized optical fields. It distinguishes between fields that have the same polarization eigentexture and conventional skyrmion number but differ in the spatial distribution of their polarized fraction. Because both \(P({\bf r})\) and, wherever \(P({\bf r})>0\), the eigentexture direction \(\hat{\bf n}_0({\bf r})\) can be reconstructed locally, the effective number can be evaluated from spatially resolved Stokes measurements.


\section*{Acknowledgments}

Research funded by the European Union NextGenerationEU/PRTR and MCIN/AEI/10.13039/501100011033 through grants \mbox{PID2022-136796OB-I00} (R.M.-H.), \mbox{PID2022-137569NBC42} (R.M.-H.), and \mbox{PID2021-127781NB-I00} (A.S.S.).


\section*{Data availability}

The data that support the findings of this article are not publicly available, as they can be reproduced from the equations provided in the work.

\vspace{-.5cm}

\appendix
\section{\label{app:complementary_profiles} Complementary radial DoP profiles}

The examples in Sec.~\ref{sec4} examine how fixed DoP profiles sample eigentextures of different winding order. Here we consider the complementary situation: the eigentexture is kept fixed while the polarized fraction is redistributed across the transverse plane.

We fix the polarization eigentexture to the \(m=1\) member of the family introduced in Sec.~\ref{sec4A} and consider the complementary radial DoP profiles
\begin{equation}
P_c(\xi)=e^{-\xi^2},
\qquad
P_o(\xi)=1-e^{-\xi^2},
\label{eq:complementary_profiles}
\end{equation}
which satisfy
\begin{equation}
P_c(\xi)+P_o(\xi)=1.
\label{eq:complementary_sum}
\end{equation}
The profile \(P_{\rm c}\) gives greater weight to the central region, whereas \(P_{\rm o}\) gives greater weight to the outer region. Although \(P_{\rm c}\) approaches zero asymptotically and \(P_{\rm o}\) vanishes at the beam center, the corresponding weighted densities have regular vanishing limits.

Using the exponential integral
\begin{equation}
E_1(x) \equiv \int_x^\infty \frac{e^{-t}}{t}\,dt,
\label{eq:exponential_integral}
\end{equation}
the corresponding effective numbers are
\begin{eqnarray}
N_{\rm eff}[P_c] & = & 2\int_0^\infty e^{-\xi^2} \frac{\xi}{(1+\xi^2)^2} \, d\xi \nonumber \\
& = & 1-e\,E_1(1) \simeq 0.4037,
\label{eq:NeffPc}
\end{eqnarray}
and
\begin{eqnarray}
N_{\rm eff}[P_o] & = & 2\int_0^\infty \left(1-e^{-\xi^2}\right) \frac{\xi}{(1+\xi^2)^2} \,d\xi \nonumber \\
& = & e\,E_1(1) \simeq 0.5963.
\label{eq:NeffPr}
\end{eqnarray}
By the linearity of \(N_{\rm eff}[P]\) with respect to the DoP profile and the complementary relation in Eq.~\eqref{eq:complementary_sum},
\begin{equation}
N_{\rm eff}[P_c] + N_{\rm eff}[P_o] = N_{\rm eff}[1] = N_1 = 1 .
\label{eq:NeffComplementarySum}
\end{equation}
The two fields therefore have the same degree-one polarization eigentexture and the same conventional skyrmion number, but different effective numbers. This comparison isolates the information added by the DoP weighting: \(N_{\rm eff}\) is sensitive to where the polarized fraction is located relative to the skyrmion-density distribution, even when the topology of the eigentexture is unchanged.



\begin{thebibliography}{21}%
\makeatletter
\providecommand \@ifxundefined [1]{%
 \@ifx{#1\undefined}
}%
\providecommand \@ifnum [1]{%
 \ifnum #1\expandafter \@firstoftwo
 \else \expandafter \@secondoftwo
 \fi
}%
\providecommand \@ifx [1]{%
 \ifx #1\expandafter \@firstoftwo
 \else \expandafter \@secondoftwo
 \fi
}%
\providecommand \natexlab [1]{#1}%
\providecommand \enquote  [1]{``#1''}%
\providecommand \bibnamefont  [1]{#1}%
\providecommand \bibfnamefont [1]{#1}%
\providecommand \citenamefont [1]{#1}%
\providecommand \href@noop [0]{\@secondoftwo}%
\providecommand \href [0]{\begingroup \@sanitize@url \@href}%
\providecommand \@href[1]{\@@startlink{#1}\@@href}%
\providecommand \@@href[1]{\endgroup#1\@@endlink}%
\providecommand \@sanitize@url [0]{\catcode `\\12\catcode `\$12\catcode
  `\&12\catcode `\#12\catcode `\^12\catcode `\_12\catcode `\%12\relax}%
\providecommand \@@startlink[1]{}%
\providecommand \@@endlink[0]{}%
\providecommand \url  [0]{\begingroup\@sanitize@url \@url }%
\providecommand \@url [1]{\endgroup\@href {#1}{\urlprefix }}%
\providecommand \urlprefix  [0]{URL }%
\providecommand \Eprint [0]{\href }%
\providecommand \doibase [0]{https://doi.org/}%
\providecommand \selectlanguage [0]{\@gobble}%
\providecommand \bibinfo  [0]{\@secondoftwo}%
\providecommand \bibfield  [0]{\@secondoftwo}%
\providecommand \translation [1]{[#1]}%
\providecommand \BibitemOpen [0]{}%
\providecommand \bibitemStop [0]{}%
\providecommand \bibitemNoStop [0]{.\EOS\space}%
\providecommand \EOS [0]{\spacefactor3000\relax}%
\providecommand \BibitemShut  [1]{\csname bibitem#1\endcsname}%
\let\auto@bib@innerbib\@empty
\bibitem [{\citenamefont {Gao}\ \emph {et~al.}(2020)\citenamefont {Gao},
  \citenamefont {Speirits}, \citenamefont {Castellucci}, \citenamefont
  {Franke-Arnold}, \citenamefont {Barnett},\ and\ \citenamefont
  {G\"otte}}]{gao:PRA:2020}%
  \BibitemOpen
  \bibfield  {author} {\bibinfo {author} {\bibfnamefont {S.}~\bibnamefont
  {Gao}}, \bibinfo {author} {\bibfnamefont {F.~C.}\ \bibnamefont {Speirits}},
  \bibinfo {author} {\bibfnamefont {F.}~\bibnamefont {Castellucci}}, \bibinfo
  {author} {\bibfnamefont {S.}~\bibnamefont {Franke-Arnold}}, \bibinfo {author}
  {\bibfnamefont {S.~M.}\ \bibnamefont {Barnett}},\ and\ \bibinfo {author}
  {\bibfnamefont {J.~B.}\ \bibnamefont {G\"otte}},\ }\bibfield  {title}
  {\bibinfo {title} {Paraxial skyrmionic beams},\ }\href
  {https://doi.org/10.1103/PhysRevA.102.053513} {\bibfield  {journal} {\bibinfo
   {journal} {Phys. Rev. A}\ }\textbf {\bibinfo {volume} {102}},\ \bibinfo
  {pages} {053513} (\bibinfo {year} {2020})}\BibitemShut {NoStop}%
\bibitem [{\citenamefont {McWilliam}\ \emph {et~al.}(2023)\citenamefont
  {McWilliam}, \citenamefont {Cisowski}, \citenamefont {Ye}, \citenamefont
  {Speirits}, \citenamefont {Götte}, \citenamefont {Barnett},\ and\
  \citenamefont {Franke-Arnold}}]{sonja:LaserPhotonRev:2023}%
  \BibitemOpen
  \bibfield  {author} {\bibinfo {author} {\bibfnamefont {A.}~\bibnamefont
  {McWilliam}}, \bibinfo {author} {\bibfnamefont {C.~M.}\ \bibnamefont
  {Cisowski}}, \bibinfo {author} {\bibfnamefont {Z.}~\bibnamefont {Ye}},
  \bibinfo {author} {\bibfnamefont {F.~C.}\ \bibnamefont {Speirits}}, \bibinfo
  {author} {\bibfnamefont {J.~B.}\ \bibnamefont {Götte}}, \bibinfo {author}
  {\bibfnamefont {S.~M.}\ \bibnamefont {Barnett}},\ and\ \bibinfo {author}
  {\bibfnamefont {S.}~\bibnamefont {Franke-Arnold}},\ }\bibfield  {title}
  {\bibinfo {title} {Topological approach of characterizing optical skyrmions
  and multi-skyrmions},\ }\href {https://doi.org/10.1002/lpor.202300155}
  {\bibfield  {journal} {\bibinfo  {journal} {Laser Photon. Rev.}\ }\textbf
  {\bibinfo {volume} {17}},\ \bibinfo {pages} {2300155} (\bibinfo {year}
  {2023})}\BibitemShut {NoStop}%
\bibitem [{\citenamefont {Shen}\ \emph {et~al.}(2024)\citenamefont {Shen},
  \citenamefont {Zhang}, \citenamefont {Shi}, \citenamefont {Du}, \citenamefont
  {Yuan},\ and\ \citenamefont {Zayats}}]{zayats:NatPhotonics:2024}%
  \BibitemOpen
  \bibfield  {author} {\bibinfo {author} {\bibfnamefont {Y.}~\bibnamefont
  {Shen}}, \bibinfo {author} {\bibfnamefont {Q.}~\bibnamefont {Zhang}},
  \bibinfo {author} {\bibfnamefont {P.}~\bibnamefont {Shi}}, \bibinfo {author}
  {\bibfnamefont {L.}~\bibnamefont {Du}}, \bibinfo {author} {\bibfnamefont
  {X.}~\bibnamefont {Yuan}},\ and\ \bibinfo {author} {\bibfnamefont {A.~V.}\
  \bibnamefont {Zayats}},\ }\bibfield  {title} {\bibinfo {title} {Optical
  skyrmions and other topological quasiparticles of light},\ }\href
  {https://doi.org/10.1038/s41566-023-01325-7} {\bibfield  {journal} {\bibinfo
  {journal} {Nat. Photon.}\ }\textbf {\bibinfo {volume} {18}},\ \bibinfo
  {pages} {15} (\bibinfo {year} {2024})}\BibitemShut {NoStop}%
\bibitem [{\citenamefont {Cheng}\ \emph {et~al.}(2026)\citenamefont {Cheng},
  \citenamefont {Rao}, \citenamefont {Ye}, \citenamefont {Zhao}, \citenamefont
  {Che}, \citenamefont {Liu}, \citenamefont {Wang},\ and\ \citenamefont
  {Shi}}]{cheng:AdvOptPhot:2025}%
  \BibitemOpen
  \bibfield  {author} {\bibinfo {author} {\bibfnamefont {C.}~\bibnamefont
  {Cheng}}, \bibinfo {author} {\bibfnamefont {L.}~\bibnamefont {Rao}}, \bibinfo
  {author} {\bibfnamefont {J.}~\bibnamefont {Ye}}, \bibinfo {author}
  {\bibfnamefont {X.}~\bibnamefont {Zhao}}, \bibinfo {author} {\bibfnamefont
  {Z.}~\bibnamefont {Che}}, \bibinfo {author} {\bibfnamefont {W.}~\bibnamefont
  {Liu}}, \bibinfo {author} {\bibfnamefont {J.}~\bibnamefont {Wang}},\ and\
  \bibinfo {author} {\bibfnamefont {L.}~\bibnamefont {Shi}},\ }\bibfield
  {title} {\bibinfo {title} {Navigating optical skyrmions---from historical
  origins to applications: tutorial},\ }\href
  {https://doi.org/10.1364/AOP.569106} {\bibfield  {journal} {\bibinfo
  {journal} {Adv. Opt. Photon.}\ }\textbf {\bibinfo {volume} {18}},\ \bibinfo
  {pages} {1} (\bibinfo {year} {2026})}\BibitemShut {NoStop}%
\bibitem [{\citenamefont {Peters}\ \emph {et~al.}(2026)\citenamefont {Peters},
  \citenamefont {Mata-Cervera}, \citenamefont {Ornelas}, \citenamefont
  {Warmbier}, \citenamefont {Shen},\ and\ \citenamefont
  {Forbes}}]{forbes2026AdvPhoton}%
  \BibitemOpen
  \bibfield  {author} {\bibinfo {author} {\bibfnamefont {C.}~\bibnamefont
  {Peters}}, \bibinfo {author} {\bibfnamefont {N.}~\bibnamefont
  {Mata-Cervera}}, \bibinfo {author} {\bibfnamefont {P.}~\bibnamefont
  {Ornelas}}, \bibinfo {author} {\bibfnamefont {R.}~\bibnamefont {Warmbier}},
  \bibinfo {author} {\bibfnamefont {Y.}~\bibnamefont {Shen}},\ and\ \bibinfo
  {author} {\bibfnamefont {A.}~\bibnamefont {Forbes}},\ }\bibfield  {title}
  {\bibinfo {title} {{Extracting topological features of skyrmionic light: a
  tutorial}},\ }\href {https://doi.org/10.1117/1.AP.8.2.023001} {\bibfield
  {journal} {\bibinfo  {journal} {Adv. Photon.}\ }\textbf {\bibinfo {volume}
  {8}},\ \bibinfo {pages} {023001} (\bibinfo {year} {2026})}\BibitemShut
  {NoStop}%
\bibitem [{\citenamefont {Mata-Cervera}, \citenamefont {Guo},\ and\ \citenamefont {Shen}}]{MataCervera2026}%
  \BibitemOpen
  \bibfield  {author} {\bibinfo {author} {\bibfnamefont {N.}~\bibnamefont
  {Mata-Cervera}}, \bibinfo {author} {\bibfnamefont {Z.}~\bibnamefont
  {Guo}},\ and\ \bibinfo {author} {\bibfnamefont {Y.}~\bibnamefont {Shen}},\ }\bibfield  {title}
  {\bibinfo {title} {{Unfolding unstable skyrmionic polarization textures}},\ }\href {https://doi.org/10.1002/lpor.71667} {\bibfield
  {journal} {\bibinfo  {journal} {Laser Photonics Rev.}},\ \bibinfo {pages} {e71667} (\bibinfo {year} {2026})}\BibitemShut
  {NoStop}%
\bibitem [{\citenamefont {Mandel}\ and\ \citenamefont
  {Wolf}(1995)}]{mandelwolf-bk}%
  \BibitemOpen
  \bibfield  {author} {\bibinfo {author} {\bibfnamefont {L.}~\bibnamefont
  {Mandel}}\ and\ \bibinfo {author} {\bibfnamefont {E.}~\bibnamefont {Wolf}},\
  }\href@noop {} {\emph {\bibinfo {title} {Optical Coherence and Quantum
  Optics}}}\ (\bibinfo  {publisher} {Cambridge University Press},\ \bibinfo
  {address} {Cambridge},\ \bibinfo {year} {1995})\BibitemShut {NoStop}%
\bibitem [{\citenamefont {Wolf}(2007)}]{wolf-bk-2}%
  \BibitemOpen
  \bibfield  {author} {\bibinfo {author} {\bibfnamefont {E.}~\bibnamefont
  {Wolf}},\ }\href@noop {} {\emph {\bibinfo {title} {Introduction to the Theory
  of Coherence and Polarization of Light}}}\ (\bibinfo  {publisher} {Cambridge
  University Press},\ \bibinfo {address} {Cambridge},\ \bibinfo {year}
  {2007})\BibitemShut {NoStop}%
\bibitem [{\citenamefont {Chen}\ \emph {et~al.}(2026)\citenamefont {Chen},
  \citenamefont {Deng}, \citenamefont {Wu}, \citenamefont {Wu}, \citenamefont
  {Liu},\ and\ \citenamefont {Hong}}]{Chen:OL:2026}%
  \BibitemOpen
  \bibfield  {author} {\bibinfo {author} {\bibfnamefont {Y.}~\bibnamefont
  {Chen}}, \bibinfo {author} {\bibfnamefont {F.}~\bibnamefont {Deng}}, \bibinfo
  {author} {\bibfnamefont {B.}~\bibnamefont {Wu}}, \bibinfo {author}
  {\bibfnamefont {R.}~\bibnamefont {Wu}}, \bibinfo {author} {\bibfnamefont
  {H.}~\bibnamefont {Liu}},\ and\ \bibinfo {author} {\bibfnamefont
  {W.}~\bibnamefont {Hong}},\ }\bibfield  {title} {\bibinfo {title} {Evolution
  of partially coherent {Stokes} skyrmions in free space},\ }\href
  {https://doi.org/10.1364/OL.600717} {\bibfield  {journal} {\bibinfo
  {journal} {Opt. Lett.}\ }\textbf {\bibinfo {volume} {51}},\ \bibinfo {pages}
  {3701} (\bibinfo {year} {2026})}\BibitemShut {NoStop}%
\bibitem [{\citenamefont {Liu}\ \emph {et~al.}(2026)\citenamefont {Liu},
  \citenamefont {Chen}, \citenamefont {Guo}, \citenamefont {Zhu}, \citenamefont
  {Chen}, \citenamefont {Cai}, \citenamefont {Shen},\ and\ \citenamefont
  {Wang}}]{Liu:arXiv:2026}%
  \BibitemOpen
  \bibfield  {author} {\bibinfo {author} {\bibfnamefont {Y.}~\bibnamefont
  {Liu}}, \bibinfo {author} {\bibfnamefont {S.}~\bibnamefont {Chen}}, \bibinfo
  {author} {\bibfnamefont {Z.}~\bibnamefont {Guo}}, \bibinfo {author}
  {\bibfnamefont {K.}~\bibnamefont {Zhu}}, \bibinfo {author} {\bibfnamefont
  {Y.}~\bibnamefont {Chen}}, \bibinfo {author} {\bibfnamefont {Y.}~\bibnamefont
  {Cai}}, \bibinfo {author} {\bibfnamefont {Y.}~\bibnamefont {Shen}},\ and\
  \bibinfo {author} {\bibfnamefont {F.}~\bibnamefont {Wang}},\ }\href
  {https://doi.org/10.48550/arXiv.2604.20207} {\bibinfo {title} {Incoherent
  light delivers skyrmionic topological resilience and transitions}} (\bibinfo
  {year} {2026}),\ \Eprint {https://arxiv.org/abs/2604.20207} {arXiv:2604.20207} \BibitemShut {NoStop}%
\bibitem [{\citenamefont {Kam}\ \emph {et~al.}(2026)\citenamefont {Kam},
  \citenamefont {Roques-Carmes}, \citenamefont {Tsesses},\ and\ \citenamefont
  {Karnieli}}]{Kam:arXiv:2026}%
  \BibitemOpen
  \bibfield  {author} {\bibinfo {author} {\bibfnamefont {A.}~\bibnamefont
  {Kam}}, \bibinfo {author} {\bibfnamefont {C.}~\bibnamefont {Roques-Carmes}},
  \bibinfo {author} {\bibfnamefont {S.}~\bibnamefont {Tsesses}},\ and\ \bibinfo
  {author} {\bibfnamefont {A.}~\bibnamefont {Karnieli}},\ }\href
  {https://doi.org/10.48550/arXiv.2604.23571} {\bibinfo {title} {Quantum
  skyrmions in mixed states of light and their nested topology}} (\bibinfo
  {year} {2026}),\ \Eprint {https://arxiv.org/abs/2604.23571} {arXiv:2604.23571} \BibitemShut {NoStop}%
\bibitem [{\citenamefont {Mart\'{i}nez-Herrero}\ and\ \citenamefont
  {Mej\'{i}as}(2007)}]{MartinezHerrero:OptLett:2007}%
  \BibitemOpen
  \bibfield  {author} {\bibinfo {author} {\bibfnamefont {R.}~\bibnamefont
  {Mart\'{i}nez-Herrero}}\ and\ \bibinfo {author} {\bibfnamefont {P.~M.}\
  \bibnamefont {Mej\'{i}as}},\ }\bibfield  {title} {\bibinfo {title} {Relation
  between degrees of coherence for electromagnetic fields},\ }\href
  {https://doi.org/10.1364/OL.32.001504} {\bibfield  {journal} {\bibinfo
  {journal} {Opt. Lett.}\ }\textbf {\bibinfo {volume} {32}},\ \bibinfo {pages}
  {1504} (\bibinfo {year} {2007})}\BibitemShut {NoStop}%
\bibitem [{\citenamefont {Schmieder}(1969)}]{schmieder:JOSA:1969}%
  \BibitemOpen
  \bibfield  {author} {\bibinfo {author} {\bibfnamefont {R.~W.}\ \bibnamefont
  {Schmieder}},\ }\bibfield  {title} {\bibinfo {title} {Stokes-algebra
  formalism},\ }\href {https://doi.org/10.1364/JOSA.59.000297} {\bibfield
  {journal} {\bibinfo  {journal} {J. Opt. Soc. Am.}\ }\textbf {\bibinfo
  {volume} {59}},\ \bibinfo {pages} {297} (\bibinfo {year} {1969})}\BibitemShut
  {NoStop}%
\bibitem [{\citenamefont {Brosseau}(1998)}]{brosseau-bk}%
  \BibitemOpen
  \bibfield  {author} {\bibinfo {author} {\bibfnamefont {C.}~\bibnamefont
  {Brosseau}},\ }\href@noop {} {\emph {\bibinfo {title} {Fundamentals of
  Polarized Light: A Statistical Optics Approach}}}\ (\bibinfo  {publisher}
  {John Wiley \& Sons},\ \bibinfo {address} {New York},\ \bibinfo {year}
  {1998})\BibitemShut {NoStop}%
\bibitem [{\citenamefont {Mart{\'\i}nez-Herrero}\ \emph
  {et~al.}(2009)\citenamefont {Mart{\'\i}nez-Herrero}, \citenamefont
  {Mej{\'\i}as},\ and\ \citenamefont {Piquero}}]{MartinezHerrero-bk}%
  \BibitemOpen
  \bibfield  {author} {\bibinfo {author} {\bibfnamefont {R.}~\bibnamefont
  {Mart{\'\i}nez-Herrero}}, \bibinfo {author} {\bibfnamefont {P.~M.}\
  \bibnamefont {Mej{\'\i}as}},\ and\ \bibinfo {author} {\bibfnamefont
  {G.}~\bibnamefont {Piquero}},\ }\href
  {https://doi.org/10.1007/978-3-642-01327-0} {\emph {\bibinfo {title}
  {Characterization of Partially Polarized Light Fields}}}\ (\bibinfo
  {publisher} {Springer},\ \bibinfo {address} {Heidelberg},\ \bibinfo {year}
  {2009})\BibitemShut {NoStop}%
\bibitem [{\citenamefont {Graybill}(1983)}]{graybill1983matrices}%
  \BibitemOpen
  \bibfield  {author} {\bibinfo {author} {\bibfnamefont {F.~A.}\ \bibnamefont
  {Graybill}},\ }\href@noop {} {\emph {\bibinfo {title} {Matrices with
  Applications in Statistics}}}\ (\bibinfo  {publisher} {Wadsworth},\ \bibinfo
  {address} {Belmont, CA},\ \bibinfo {year} {1983})\BibitemShut {NoStop}%
\bibitem [{\citenamefont {Han}(2017)}]{han-bk}%
  \BibitemOpen
  \bibfield  {author} {\bibinfo {author} {\bibfnamefont {J.~H.}\ \bibnamefont
  {Han}},\ }\href {https://doi.org/10.1007/978-3-319-69246-3} {\emph {\bibinfo
  {title} {Skyrmions in Condensed Matter}}}\ (\bibinfo  {publisher}
  {Springer},\ \bibinfo {address} {Cham, Switzerland},\ \bibinfo {year}
  {2017})\BibitemShut {NoStop}%
\bibitem [{\citenamefont {Belavin}\ and\ \citenamefont
  {Polyakov}(1975)}]{belavin:JETPLett:1975}%
  \BibitemOpen
  \bibfield  {author} {\bibinfo {author} {\bibfnamefont {A.~A.}\ \bibnamefont
  {Belavin}}\ and\ \bibinfo {author} {\bibfnamefont {A.~M.}\ \bibnamefont
  {Polyakov}},\ }\bibfield  {title} {\bibinfo {title} {Metastable states of
  two-dimensional isotropic ferromagnets},\ }\href
  {http://jetpletters.ru/ps/0/article_23383.shtml} {\bibfield  {journal}
  {\bibinfo  {journal} {JETP Lett.}\ }\textbf {\bibinfo {volume} {22}},\
  \bibinfo {pages} {245} (\bibinfo {year} {1975})}\BibitemShut {NoStop}%
\bibitem [{\citenamefont {Ye}\ \emph {et~al.}(2024)\citenamefont {Ye},
  \citenamefont {Barnett}, \citenamefont {Franke-Arnold}, \citenamefont
  {G\"otte}, \citenamefont {McWilliam}, \citenamefont {Speirits},\ and\
  \citenamefont {Cisowski}}]{sonja:PRSA:2024}%
  \BibitemOpen
  \bibfield  {author} {\bibinfo {author} {\bibfnamefont {Z.}~\bibnamefont
  {Ye}}, \bibinfo {author} {\bibfnamefont {S.~M.}\ \bibnamefont {Barnett}},
  \bibinfo {author} {\bibfnamefont {S.}~\bibnamefont {Franke-Arnold}}, \bibinfo
  {author} {\bibfnamefont {J.~B.}\ \bibnamefont {G\"otte}}, \bibinfo {author}
  {\bibfnamefont {A.}~\bibnamefont {McWilliam}}, \bibinfo {author}
  {\bibfnamefont {F.~C.}\ \bibnamefont {Speirits}},\ and\ \bibinfo {author}
  {\bibfnamefont {C.~M.}\ \bibnamefont {Cisowski}},\ }\bibfield  {title}
  {\bibinfo {title} {Theory of paraxial optical skyrmions},\ }\href
  {https://doi.org/10.1098/rspa.2024.0109} {\bibfield  {journal} {\bibinfo
  {journal} {Proc. R. Soc. A}\ }\textbf {\bibinfo {volume} {480}},\ \bibinfo
  {pages} {20240109} (\bibinfo {year} {2024})}\BibitemShut {NoStop}%
\bibitem [{\citenamefont {Sj\"oqvist}\ \emph {et~al.}(2000)\citenamefont
  {Sj\"oqvist}, \citenamefont {Pati}, \citenamefont {Ekert}, \citenamefont
  {Anandan}, \citenamefont {Ericsson}, \citenamefont {Oi},\ and\ \citenamefont
  {Vedral}}]{sjoqvist:PRL:2000}%
  \BibitemOpen
  \bibfield  {author} {\bibinfo {author} {\bibfnamefont {E.}~\bibnamefont
  {Sj\"oqvist}}, \bibinfo {author} {\bibfnamefont {A.~K.}\ \bibnamefont
  {Pati}}, \bibinfo {author} {\bibfnamefont {A.}~\bibnamefont {Ekert}},
  \bibinfo {author} {\bibfnamefont {J.~S.}\ \bibnamefont {Anandan}}, \bibinfo
  {author} {\bibfnamefont {M.}~\bibnamefont {Ericsson}}, \bibinfo {author}
  {\bibfnamefont {D.~K.~L.}\ \bibnamefont {Oi}},\ and\ \bibinfo {author}
  {\bibfnamefont {V.}~\bibnamefont {Vedral}},\ }\bibfield  {title} {\bibinfo
  {title} {Geometric phases for mixed states in interferometry},\ }\href
  {https://doi.org/10.1103/PhysRevLett.85.2845} {\bibfield  {journal} {\bibinfo
   {journal} {Phys. Rev. Lett.}\ }\textbf {\bibinfo {volume} {85}},\ \bibinfo
  {pages} {2845} (\bibinfo {year} {2000})}\BibitemShut {NoStop}%
\bibitem [{\citenamefont {Uhlmann}(1986)}]{uhlmann1986parallel}%
  \BibitemOpen
  \bibfield  {author} {\bibinfo {author} {\bibfnamefont {A.}~\bibnamefont
  {Uhlmann}},\ }\bibfield  {title} {\bibinfo {title} {Parallel transport and
  ``quantum holonomy'' along density operators},\ }\href
  {https://doi.org/10.1016/0034-4877(86)90055-8} {\bibfield  {journal}
  {\bibinfo  {journal} {Rep. Math. Phys.}\ }\textbf {\bibinfo {volume} {24}},\
  \bibinfo {pages} {229} (\bibinfo {year} {1986})}\BibitemShut {NoStop}%
\bibitem [{\citenamefont {Cisowski}\ \emph {et~al.}(2023)\citenamefont
  {Cisowski}, \citenamefont {Ross},\ and\ \citenamefont
  {Franke-Arnold}}]{franke-arnold:AdvPhotonRes:2023}%
  \BibitemOpen
  \bibfield  {author} {\bibinfo {author} {\bibfnamefont {C.}~\bibnamefont
  {Cisowski}}, \bibinfo {author} {\bibfnamefont {C.}~\bibnamefont {Ross}},\
  and\ \bibinfo {author} {\bibfnamefont {S.}~\bibnamefont {Franke-Arnold}},\
  }\bibfield  {title} {\bibinfo {title} {Building paraxial optical skyrmions
  using rational maps},\ }\href {https://doi.org/10.1002/adpr.202200350}
  {\bibfield  {journal} {\bibinfo  {journal} {Adv. Photon. Res.}\ }\textbf
  {\bibinfo {volume} {4}},\ \bibinfo {pages} {2200350} (\bibinfo {year}
  {2023})} \BibitemShut {NoStop}%
\end{thebibliography}



%

\end{document}